\documentclass[preprint]{iucr}
\RequirePackage{graphicx}
\RequirePackage{upgreek}
\RequirePackage{dcolumn}   
\RequirePackage{bm}        
\RequirePackage{amssymb}   
\RequirePackage[english]{babel}
\RequirePackage[utf8x]{inputenc}
\RequirePackage{subfigure}
\RequirePackage{textcomp}
\RequirePackage{eso-pic}
\RequirePackage{wasysym}
\RequirePackage{fancyhdr}
\RequirePackage{lmodern}
\RequirePackage{siunitx}

\journalcode{S} 

\begin{document}

\title{Anisotropy enhanced X-ray scattering from solvated transition metal complexes}

\cauthor[a,b]{Elisa}{Biasin}{elbia@fysiks.dtu.dk}{}
\author[a,f]{Tim B.}{van Driel}
\author[c]{Gianluca}{Levi}
\author[a]{Mads G.}{Laursen}
\author[d]{Asmus O.}{Dohn}
\author[a]{Asbj\o rn}{Moltke}
\author[a]{Peter}{Vester}
\author[a]{Frederik B. K.}{Hansen}
\author[a,b,e]{Kasper S.}{Kj\ae r}
\author[a]{Tobias}{Harlang}
\author[b]{Robert}{Hartsock}
\author[a]{Morten}{Christensen}
\author[b]{Kelly J.}{Gaffney}
\author[c]{Niels E.}{Henriksen}
\author[c]{Klaus B.}{M\o ller}
\author[a]{Kristoffer}{Haldrup}
\author[a]{Martin M.}{Nielsen}

\aff[a]{Department of Physics, Technical University of Denmark, Fysikvej 307, DK-2800 Kongens Lyngby, \country{Denmark}}
\aff[b]{PULSE Institute, SLAC National Accelerator Laboratory, Menlo Park, California 94025, USA}
\aff[c]{Department of Chemistry, Technical University of Denmark, Kemitorvet 207, DK-2800 Kongens Lyngby, Denmark}
\aff[d]{Faculty of Physical Sciences, University of Iceland}
\aff[e]{Department of Chemical Physics, Lund University, Box 118, S-22100 Lund, Sweden}
\aff[f]{Linac Coherent Light Source, SLAC National Accelerator Laboratory, Menlo Park, California 94025, \country{USA}}


\begin{abstract}
Time-resolved X-ray scattering patterns from photoexcited molecules in solution are in many cases anisotropic at the ultrafast time scales accessible at X-ray Free Electron Lasers (XFELs). This anisotropy arises from the interaction of a linearly polarized UV-vis pump laser pulse with the sample, which induces anisotropic structural changes that can be captured by femtosecond X-ray pulses. In this work we describe a method for quantitative analysis of the anisotropic scattering signal arising from an ensemble of molecules and we demonstrate how its use can enhance the structural sensitivity of the time-resolved X-ray scattering experiment. We apply this method on time-resolved X-ray scattering patterns measured upon photoexcitation of a solvated di-platinum complex at an XFEL and explore the key parameters involved. We show that a combined analysis of the anisotropic and isotropic difference scattering signals in this experiment allows a more precise determination of the main photoinduced structural change in the solute, i.e. the change in Pt-Pt bond length, and yields more information on the excitation channels than the analysis of the isotropic scattering only.  Finally, we discuss how the anisotropic transient response of the solvent can enable the determination of key experimental parameters such as the Instrument Response Function. 
\end{abstract}


\maketitle

\section{Introduction}
Time-resolved X-ray diffuse scattering (XDS) experiments give insight into the photoinduced structural dynamics of solvated molecules. In these experiments, a laser pulse initiates the dynamic process, which is subsequently probed by an X-ray probe pulse arriving at specific time delays after the pump event. If the laser pulse is ultra-short, the ensuing structural dynamics are coherently initiated in the molecular ensemble~\cite{zewail}, and the scattering signal can be used to retrieve structural changes occurring in the molecule after the electronic excitation~\cite{Borfecchia2013,Biasin2016,Kreuff-Fe,AdachiNature2015,TimIrDimer,CherguiReview2017,kong2008}. Such experiments are of fundamental importance for understanding the structure-function relationship of, for instance, transition metal complexes whose photochemical and photophysical properties can be applied in technologies such as solar energy conversion and photocatalysis~\cite{AbundantMetalsforCo,hydrogen,belloTM}. 

\begin{figure}
\includegraphics[scale=0.5,bb=0 0 500 120]{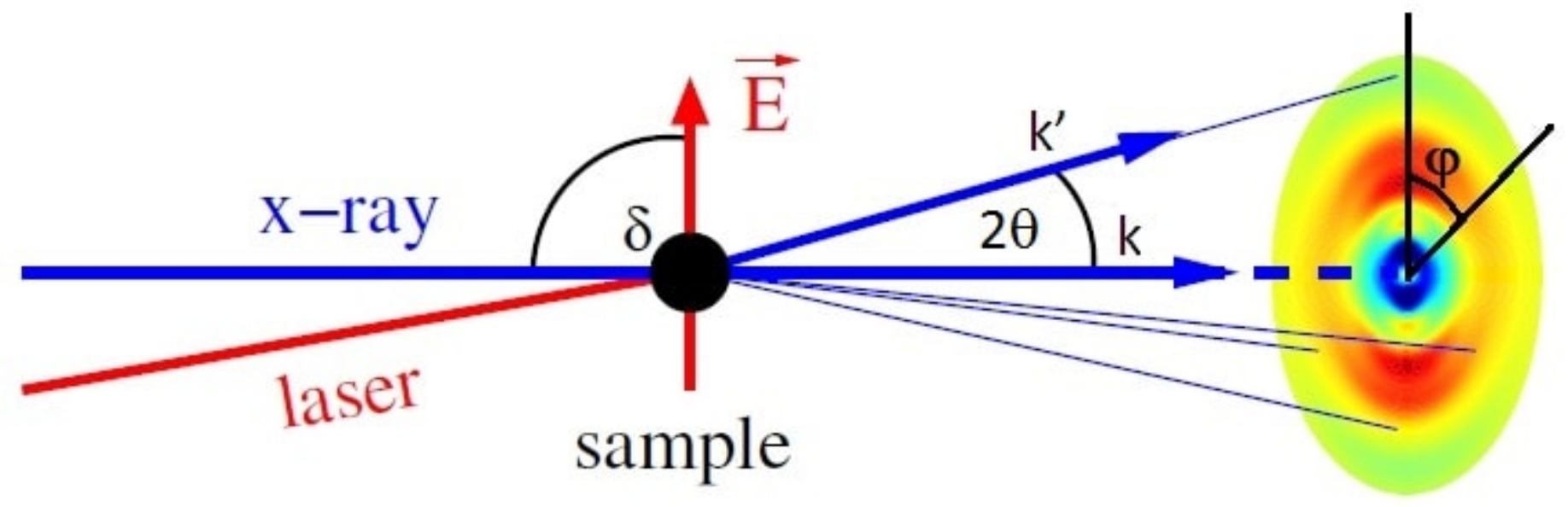}
\caption{\small Experimental setup for a standard time-resolved XDS experiment. $\delta$ is the angle between the direction of propagation of the X-ray beam and the laser polarization axis  \textbf{E}; 2$\theta$ is the scattering angle, i.e. the angle between the incoming (\textbf{k}) and outgoing (\textbf{k´}) momentum of the X-ray beam; $\varphi$ is the angle between the projection of the laser polarization and the scattering vector \textbf{Q} = \textbf{k´} - \textbf{k} on the detector surface.}
\label{ssetup}
\end{figure} 

In typical time-resolved XDS experiments, laser pump and X-ray probe pulses are focused on a thin liquid jet, which is produced by pumping the liquid sample through a nozzle. The diffuse X-ray scattering is collected on a two dimensional (2D) detector placed after the sample on the plane perpendicular to the propagation direction of the X-ray beam  (Fig.~\ref{ssetup}). The measured scattering signal contains information about all the inter-nuclear distances of the sample at a specific time delay and it is usually  dominated by the scattering from the solvent. The sensitivity to the solute is enhanced in the difference scattering signal, which is constructed by  subtracting the signal collected without photoexciting the sample from the signal collected after photoexcitation. In this way the unchanging background contributions cancel out, and the difference scattering signal arises from the changes in the inter-atomic distances in the probed sample volume~\cite{Borfecchia2013,IheeReview2010}. The established procedure for analysing scattering data from liquid samples consists of an azimuthal integration of the 2D difference scattering patterns, since the signal is usually assumed to arise from an isotropically distributed ensemble of molecules~\cite{IheeReview2010,haldrup2010analysis}. If the sample is isotropic, the azimuthal integration allows for an improvement of signal-to-noise (S/N) without loss of information.  This procedure was first established with synchrotron data and it is justified when the time resolution is longer than the molecular rotational correlation time in solution, which ranges in the 10-100 ps time scale for transition metal complexes~\cite{FSbook,kim2015gold}.

\begin{figure}
\begin{center}
\includegraphics[scale=0.2,bb=0 0 1800 500]{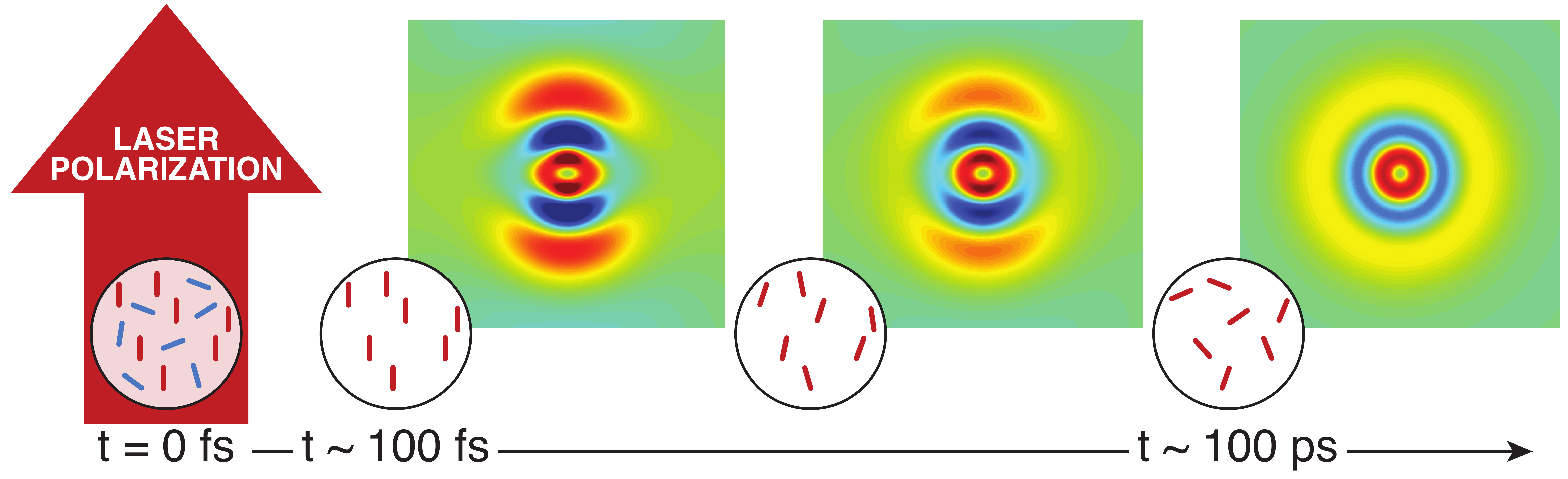}
\end{center}
\caption{\small Illustration of the photoselection process. Before the arrival of the laser pump ($t < 0$), the molecules are randomly oriented in solution. The laser pulse  preferentially excites the molecules with the transition dipole moments parallel to the laser polarization axis. Thus, the excited state population is created as an aligned ensemble (at t = 0 the distribution of the transition dipole moment in molecules being excited is a cosine-squared distribution with respect to the laser polarization axis) yielding anisotropic difference scattering patterns. After the excitation event ($t > 0$), the rotation  of both the excited and unexcited populations causes rotational dephasing of the alignment.}
\label{photoSel}
\end{figure}

At X-ray Free Electron Lasers (XFELs), that can deliver femtosecond X-ray pulses, the photoinduced structural changes in the sample can be captured at the time scale of atomic motions. This enables the observation of vibrational and rotational molecular dynamics in, for instance, solvated transition metal complexes~\cite{Biasin2016,kim2015gold,Lemke2015,CherguiReview2017}. If the photoselection process (illustrated in Fig.~\ref{photoSel} and further described below) creates an aligned excited state ensemble of molecules,  anisotropic 2D scattering patterns can be observed on time scales shorter than the rotational correlation time of the molecules~\cite{kim2011protein,kim2015gold,Yang2016,YangPRL2016,glownia}. Anisotropic scattering patterns contain information on the molecular preferred orientation, and thus can add sensitivity to spatial degrees of freedom compared to isotropic patterns~\cite{Centurion2012,Burger2010,PhysRevLett112}.  The theoretical foundation for  interpreting anisotropic scattering contributions from aligned ensembles of molecules in time-resolved experiments has been laid out in some detail over the past ten years~\cite{baskin&zewaill_2006,aniUlf,Hub_2015,Penfold_ANI}. However, quantitative structural analysis of anisotropic scattering data from molecular ensembles where a photoselected sub-population has been promoted to an electronic excited state has been demonstrated only in a very few cases in gas-phase~\cite{Yang2016,YangPRL2016,glownia}, and a robust methodology for such cannot yet be said to have been fully established as the debate surrounding the work of Glownia $at$ $al.$ shows ~\cite{MukamelComm,GlowniaReply}.  Specifically with respect to solution-state molecular systems, only qualitative analysis of time-resolved anisotropic scattering data have been reported, and the anisotropic contributions to the observed scattering patterns have been mostly used only to obtain the rotational correlation time of the molecules in solution~\cite{kim2011protein,kim2015gold}. 

In this article, we revisit the formalism derived by Lorenz $et$ $al$.~\cite{aniUlf} for analysing scattering from a photo-aligned ensemble of molecules. We use this formalism to extract the isotropic and anisotropic contribution from 2D scattering patterns measured at an XFEL upon photoexcitation of a transition metal complex in solution and we explore a few of the key parameters involved in this procedure. We show that, in the presence of anisotropic scattering, the separation of the two contributions can, and should, replace the standard azimuthal integration in the data reduction procedures. As a key point, we demonstrate that the anisotropic contribution can be quantitatively analysed in order to extract structural information about the photoexcited ensemble of molecules. These results are discussed in the framework of how the information extracted from a combined analysis of the isotropic and anisotropic contributions to the measured difference scattering signals can potentially help disentangling the many inter- and intra- nuclear degrees of freedom involved in photoinduced structural dynamics of molecules in solution.

\begin{figure}
\begin{center}
\includegraphics[scale=0.29,bb=0 0 650 550]{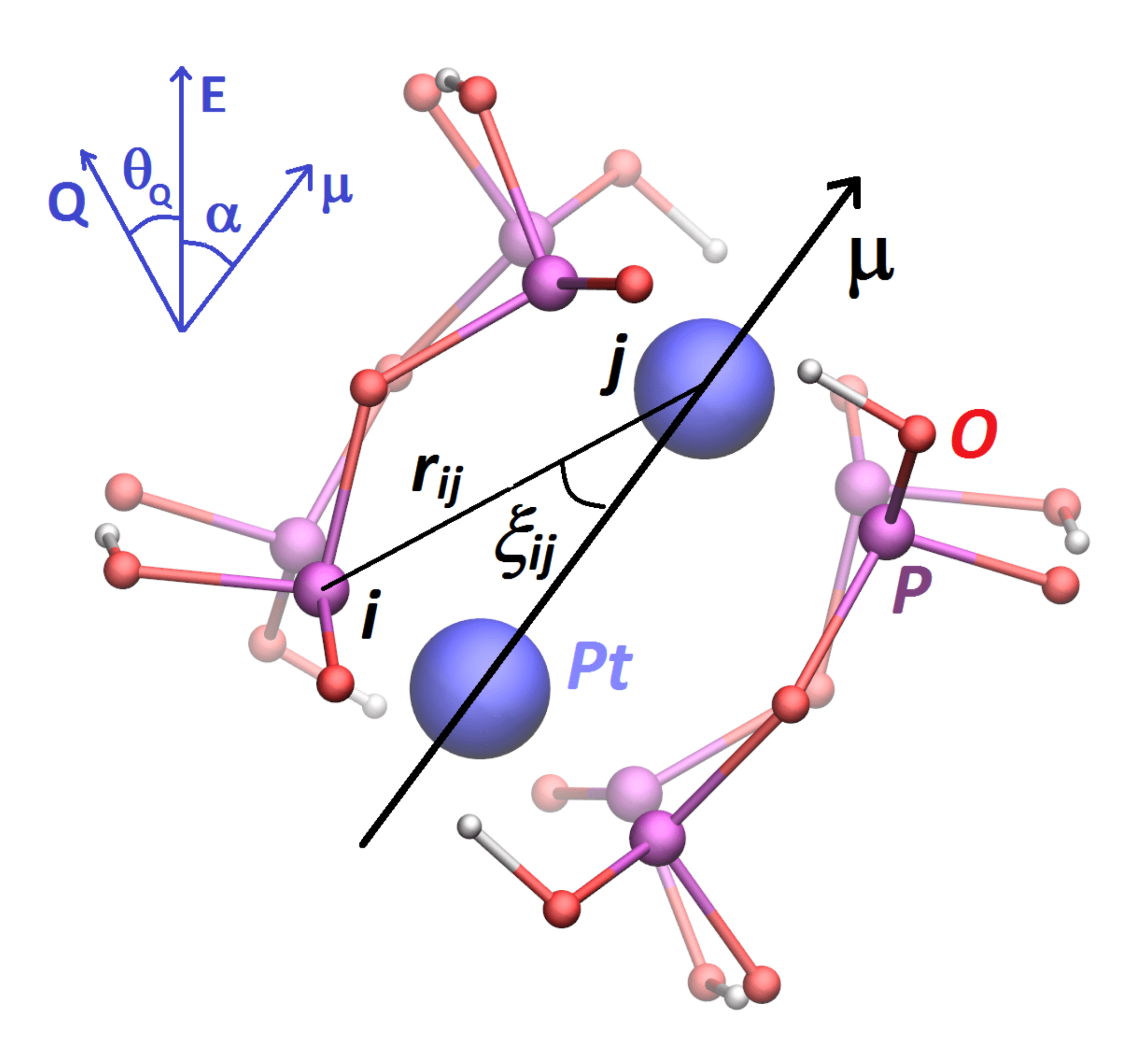}
\caption{\small Molecular structure of PtPOP and relevant angles. PtPOP is a symmetric top molecule with the main axis of symmetry parallel to the Pt-Pt axis~\cite{zipp,GRAY2017297}.  In the molecular fixed frame (black arrows): $\xi_{ij}$  is the angle between the internuclear displacement {\boldmath$r_{ij}$} and the transition dipole moment {\boldmath$\upmu$} of the molecule, which is parallel to the main axis of symmetry of the molecule. In the laboratory frame (blue arrows): $\alpha$ is the angle between the laser polarization axis \textbf{E} and the transition dipole vector {\boldmath$\upmu$} of the molecule, $\theta_{Q}$ is the angle between the laser polarization axis and the scattering vector.}
\label{simTop}
\end{center}
\end{figure}

The method is exemplified on scattering data collected at the X-ray Pump Probe (XPP) instrument~\cite{xpp} at the Linac Coherent Light Source upon photoexcitation of the  tetrakis-$\upmu$-pyrophosphitodiplatinate(II) ion [Pt$_{2}$(P$_{2}$O$_{5}$H$_{2}$)$_{4}]^{4-}$, here abbreviated PtPOP. Fig.~\ref{simTop} shows the molecular structure of PtPOP: a Pt-Pt dimer is held together by four pyrophosphito ligands. The compound belongs to the C4h point group, with approximately four-fold symmetry along the Pt-Pt axis~\cite{zipp,GRAY2017297}. Thus, the Pt-Pt axis has higher rotational symmetry than the other two axes, and the molecule is a symmetric top (i.e. two principal moments of inertia have the same value and the third has a unique value).  This compound was first synthesized in 1977 and has been object of a vast amount of studies, due to its characteristic photophysical properties and activity as a photocatalyst~\cite{Roundhill1989,Stiegman1987}. In the last decades PtPOP has  become a model compound for time-resolved X-ray studies, due to its high scattering power, its long singlet excited state lifetime, and its high symmetry~\cite{PtPOP2009,vanDerVeen2009}.  For this experiment, an 80 mM aqueous solution of PtPOP was circulated through a nozzle producing a 50 $\upmu$m round liquid jet. Each photocycle was initiated by a 3 $\upmu$J laser pulse at 395 nm and with 50 fs pulse width (FWHM), focused onto a 150 $\upmu$m diameter spot. The scattering from the 9.5 keV X-ray probe pulses were detected by the Cornell-SLAC Pixel Array Detector (CSPAD)~\cite{cspad} covering scattering vectors up to 4.5 \AA{}$^{-1}$. The time delay \textit{t} between the laser and the X-ray pulses was determined for every pump-probe event with $\sim$10 fs (FWHM) resolution using the XPP timing-tool~\cite{timingtoolXPP}. The detector signal was corrected according to the procedure described by van Driel~\cite{timSVD}, including corrections for the detector geometry and the horizontal polarization of the X-ray beam. The corrected scattering signal was scaled to the liquid unit cell reflecting the stoichiometry of the sample~\cite{haldrup2010analysis}, yielding the acquired signal in electron units per solute molecule (e.u./molec.). Individual 2D difference scattering patterns were then time-sorted and averaged. Movie~S1 of the Supporting Online Information (SOI) shows averaged measured difference scattering patterns upon photoexcitation of PtPOP as a function of increasing time delay: a strongly anisotropic difference scattering signal is visible at early time delay and decays on a $\sim$100 ps time scale. The signal is expected to arise predominantly from the shortening of the Pt-Pt distance since the 395 nm transition, which has polarization along the Pt-Pt bond, involves the promotion of an electron from the antibonding d$\sigma^{*}$ to the bonding p$\sigma$ orbital~\cite{Stiegman1987}. Christensen $et$ $al$. investigated the structure of the excited state of PtPOP in water with time-resolved XDS and obtained a Pt-Pt contraction of 0.24 $\pm$ 0.06 \AA{} with respect to the ground state of the molecule~\cite{PtPOP2009}. Van der Veen $et$ $al.$ concluded a 0.31 $\pm$ 0.05 \AA{} Pt-Pt contraction upon excitation of the complex in ethanol, with X-ray absorption spectroscopy~\cite{vanDerVeen2009}. The well-defined transition dipole moment and the strong scattering signal arising from the contraction of the Pt-Pt bond makes PtPOP an ideal model system to benchmark the formalism required in order to interpret anisotropic difference scattering patterns.

\section{Methods}
In time-resolved XDS experiments, the molecules in solution are randomly oriented before the arrival of the laser pump. If the linearly polarized ultrashort laser pulse interacts with a single transition dipole moment {\boldmath$\upmu$} of the molecule, the angular distribution $D$ of this transition dipole moment in molecules being excited displays cylindrical symmetry with respect to the laser polarization  \textbf{E}.
Therefore, this distribution can be expressed through an expansion in Legendre polynomials $P_{n} (\cos\alpha)$  
of only the angle $\alpha$ between the laser polarization axis and the transition dipole moment~\cite{baskin&zewaill_2006}:
\begin{equation}
D(\alpha, \eta)= D(\alpha) = \sum_{n = 0}^{\infty} c_{n}P_{n}(\cos\alpha)
\label{eq_easy}
\end{equation}
where $\alpha$ and $\eta$ are the polar and azimuthal angles, respectively, in a spherical coordinates system with polar axis \textbf{E}, and $c_{n}$ are expansion coefficients. In case of one-photon absorption of linearly polarized radiation by an isotropic molecular ensemble in thermal equilibrium, the distribution of the transition dipole moment in molecules being excited is a cosine-squared distribution with respect to the laser polarization axis  ~\cite{Moller,baskin&zewaill_2006}. Cast into the generic form of Eq.~\ref{eq_easy}, the distribution of excited state molecules immediately after the excitation event can be expressed as a sum of two terms:
\begin{equation}
D (\alpha) = c_{0}P_{0}(\cos\alpha) + c_{2}P_{2}(\cos\alpha)
\label{secondEq}
\end{equation}
where $P_{0}(x) = 1$ and $P_{2}(x) = \frac{1}{2}(3x^{2} - 1)$, with $c_{2}$ = 2$c_{0}$. 
Properly normalizing~\cite{SimmTopAlignment} the angular distribution in Eq.~\ref{eq_easy} on the unit sphere gives:
\begin{equation}
1=\frac{1}{2} \int_{0}^{\pi} D(\alpha) \sin\alpha d\alpha = c_{0}
\label{uff}
\end{equation}
where the last equality is a consequence of the orthogonality of the Legendre polynomials. 
 
Assuming ultrashort laser pump and X-ray probe pulses, the differential scattering cross-section from an ensemble of excited molecules, at a specific pump-probe delay time $t$, can be expressed by weighting the squared molecular form factor with the instantaneous distribution $\rho$ of nuclear geometries~\cite{aniUlf}: 
\begin{equation}
\frac{d\sigma}{d\Omega}(\textbf{Q}, t) = \sigma_{T} \int d \textbf{R}  \rho (\textbf{R}, t) |F_{mol}(\textbf{Q}, \textbf{R})|^{2}
\label{general}
\end{equation}
where $\textbf{R}$ has dimensions 3$N$ and describes the nuclear coordinates in the laboratory frame, $\sigma_{T}$ is the Thomson scattering cross-section and $\textbf{Q}$ is the scattering vector. After assuming that rotational and vibrational degrees of freedom of the molecule are uncoupled~\cite{aniUlf}, we express the rotational part of the distribution through $D(\alpha)$ in Eq.~\ref{eq_easy} and the vibrational distribution as a $\delta$-distribution. Under these assumptions, the scattering from an ensemble of molecules with their transition dipole moments aligned with respect to the laser polarization axis can be calculated through:
\begin{equation}
\frac{d\sigma}{d\Omega}(\textbf{Q}, t)  = 2\pi \sigma_{T} \int_{0}^{\pi} d \alpha \sin\alpha D (\alpha, t) |F_{mol}(\textbf{Q},  \alpha, \textbf{r}(t))|^{2},
\label{lessgeneral}
\end{equation}
where $\textbf{r}$ is the vector of inter-nuclear distances. Furthermore, this expression (Eq.~\ref{lessgeneral}) can be simplified by utilizing that the molecule is a symmetric top  with the transition dipole moment parallel to the unique axis of symmetry. Under this assumption, as detailed in Baskin and Zewail~\cite{baskin&zewaill_2005}, the angular distribution of each intra-molecular distance {\boldmath$r_{ij}$} with respect to the laser polarization axis can be expressed through the angle $\xi_{ij}$  between {\boldmath$r_{ij}$} and the main axis of symmetry of the molecule. These parameters are illustrated in Fig.~\ref{simTop}: PtPOP is a symmetric top molecule with the main axis of symmetry parallel to the Pt-Pt bond, which is also the direction of the transition dipole moment photoselected in this experiment and therefore meets the above assumptions. As described by Lorenz et al.~\cite{aniUlf}, under the described assumptions, the integral in  Eq.~\ref{lessgeneral} has the analytical form:
\begin{equation}
\frac{d\sigma}{d\Omega}(Q, \theta_{Q}, t)  = 2 (2 \pi)^{2} \sigma_{T} \sum_{n}  P_{n} (\cos\theta_{Q}) S_{n}(Q, t)
\label{all}
\end{equation}
where $\theta_{Q}$ is the angle between the laser polarization axis and \textbf{Q} (see Fig.~\ref{simTop}), and
\begin{equation}
S_{n}(Q, t) = (-1)^{n/2} c_{n}(t) \sum_{i,j}^{N}f_{i}(Q)f_{j}(Q) P_{n} (\cos\xi_{ij}(t))j_{n}(Qr_{ij}(t))
\end{equation}
where $j_{n}$ are the spherical Bessel functions and $f_{i}$ is the atomic form factor of atom $i$, which is used to express the molecular form factor within the Independent Atom Model~\cite{Moller}. 
Assuming one-photon absorption (and thus the angular distribution in Eq.~\ref{secondEq}), only the $n = 0$ and $n = 2$ terms contribute in Eq.~\ref{all}~\cite{aniUlf} and the scattering signal can be written as:
\begin{equation}
\frac{d\sigma}{d\Omega}(Q, \theta_{Q},t) \propto  S_{0}(Q,t) +  P_{2}(\cos\theta_{Q})S_{2}(Q,t).
\label{mainEq}
\end{equation}
Since $j_{0}(x) = \frac{\sin x}{x}$, $S_{0}$ is recognized as the Debye formula for isotropic ensembles:
\begin{equation}
S_{0}(Q,t) = \sum_{i,j}^{N}f_{i}(Q)f_{j}(Q) \frac{\sin(Qr_{ij}(t))}{Qr_{ij}(t)},
\label{mains0}
\end{equation}
while $S_{2}$ contains information about the orientation of the single bond {\boldmath$r_{ij}$} with respect to the transition dipole moment of the molecule:
\begin{equation}
S_{2}(Q,t) = - c_{2}(t) \sum_{i,j}^{N}f_{i}(Q)f_{j}(Q)P_{2} (\cos\xi_{ij}(t))j_{2}(Qr_{ij}(t)),
\label{mains2}
\end{equation} 
with $j_{2}(x)= (\frac{3}{x^{2}} - 1) \frac{\sin x}{x} - 3\frac{\cos x}{x^{2}}$.
From here $S_{0}$ and $S_{2}$ will be referred to as the isotropic and anisotropic part of the scattering signal, respectively. $c_{2}$ decays from its initial value of 2 to zero according to the rotational correlation time of the molecules in solution.

The excitation will leave a hole in the ground state distribution of the same rotational anisotropy as the excited-state ensemble~\cite{baskin&zewaill_2006,fleming}. Starting from Eq.~\ref{mainEq}, the difference scattering signal $\Delta S$ including both excited-state and ground-state contributions can, therefore, be decomposed as~\cite{aniUlf}: 
\begin{equation}
\Delta S(Q, \theta_{Q}, t) \propto \Delta S_{0}(Q, t) + P_{2}(\cos\theta_{Q}) \Delta S_{2}(Q, t)
\label{allDiff}
\end{equation}
where $\Delta S_{0}(Q, t)$ and $\Delta S_{2}(Q, t)$ are, respectively, the isotropic and anisotropic difference scattering signals arising from the photoinduced structural changes in the sample at a specific pump-probe time delay \textit{t}. 

With respect to  Eq.~\ref{allDiff}, $S_{0}$ and $S_{2}$ depend only on internal coordinates of the molecule, while $P_{2}(\cos\theta_{Q})$ depends only on the geometry of the experiment. Specifically, as detailed in~\cite{baskin&zewaill_2006}, $\cos\theta_{Q}$ can be expressed as a function of the angles $\delta$, $\theta$ and $\varphi$, as they are defined in Fig.~\ref{ssetup}:
\begin{equation}
\cos\theta_{Q} = \sin \theta \cos\delta - \cos\theta \cos\varphi \sin\delta.
\label{costhetaq}
\end{equation}
In the case of most standard XDS experiments, and for the PtPOP experiment presented here, the laser beam is nearly collinear to the X-ray beam. This is one of the possible configurations that yields a 90 degrees angle between the laser polarization axis and the direction of propagation of the X-ray beam ($\delta$ in Fig.~\ref{ssetup}), and Eq.~\ref{costhetaq} simplifies to: 
\begin{equation}
\cos\theta_{Q} =  - \cos\theta \cos\varphi.
\label{costhetaq_final}
\end{equation}
Since $\varphi$ is the azimuthal angle on the detector surface, $\cos \theta_{Q}$ maps $P_{2}$  anisotropically onto the detector surface. The second order Legendre polynomial, and hence the anisotropic contribution to the scattering, vanishes when $\theta_{Q}$ in Eq.~\ref{costhetaq} is equal to the magic angle. However, as extensively described by Baskin and Zewail~\cite{baskin&zewaill_2006}, it is not possible to simultaneously remove anisotropic contributions from the entire scattering pattern for a given choice of $\delta$. The linear relation between scattering intensity and $P_{2}$ in Eq.~\ref{allDiff} can be used to retrieve, for a specific \textit{Q} value, the isotropic and anisotropic difference scattering signal, as described in following.

\begin{figure}
\includegraphics[scale=0.3,bb=0 0 600 600]{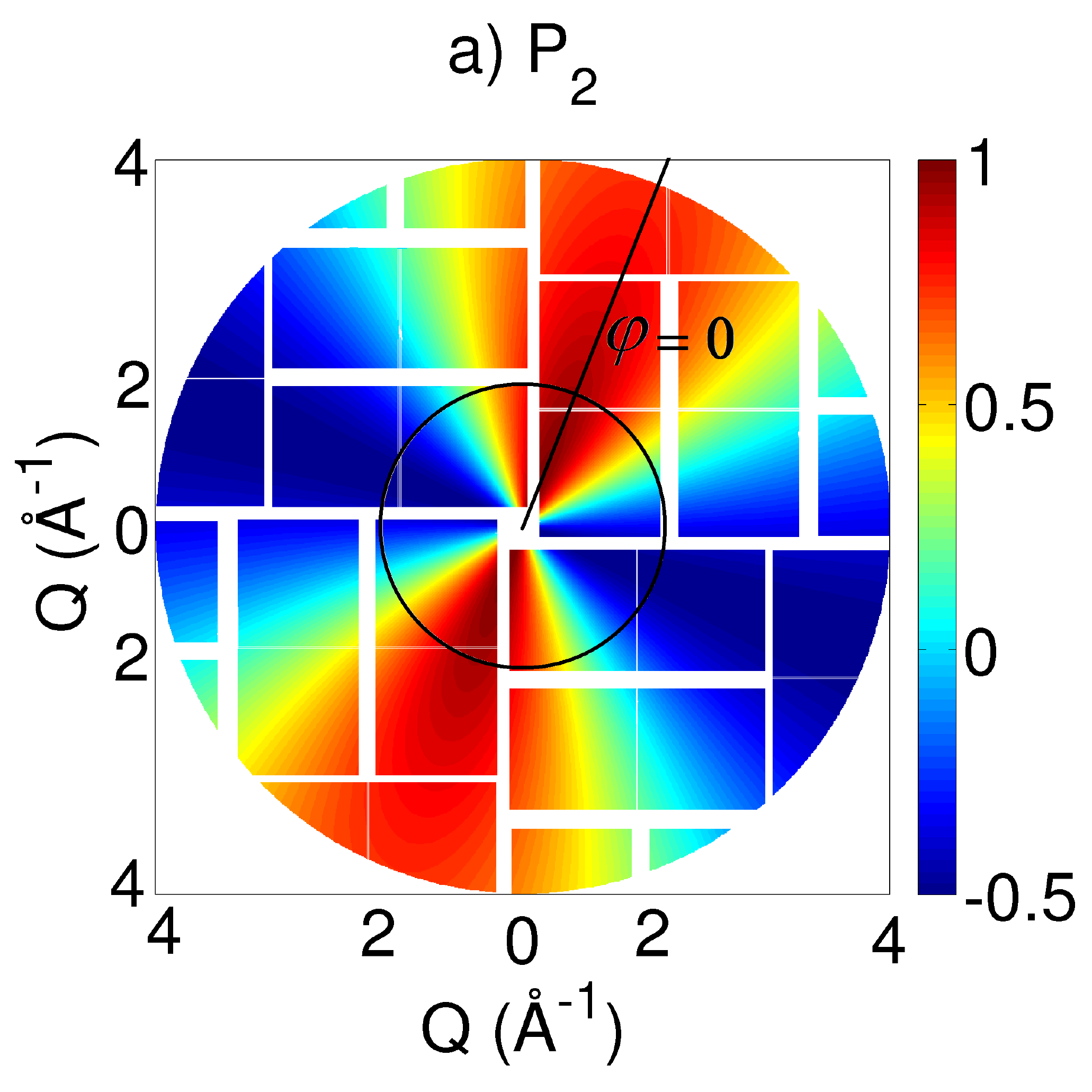}
\includegraphics[scale=0.3,bb=0 0 600 600]{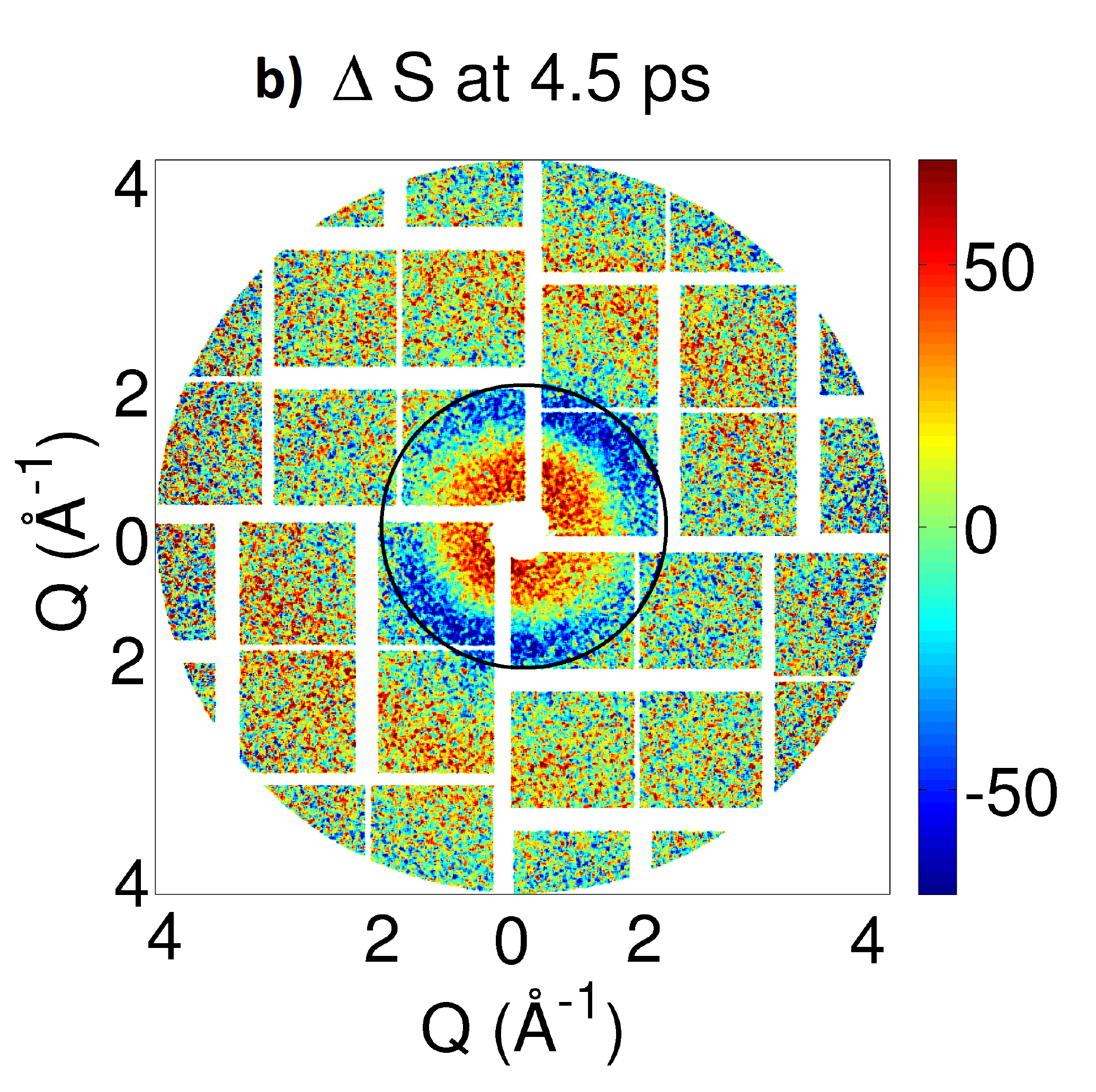}
 \makebox[1\linewidth]{
\includegraphics[scale=0.25,bb=0 0 800 600]{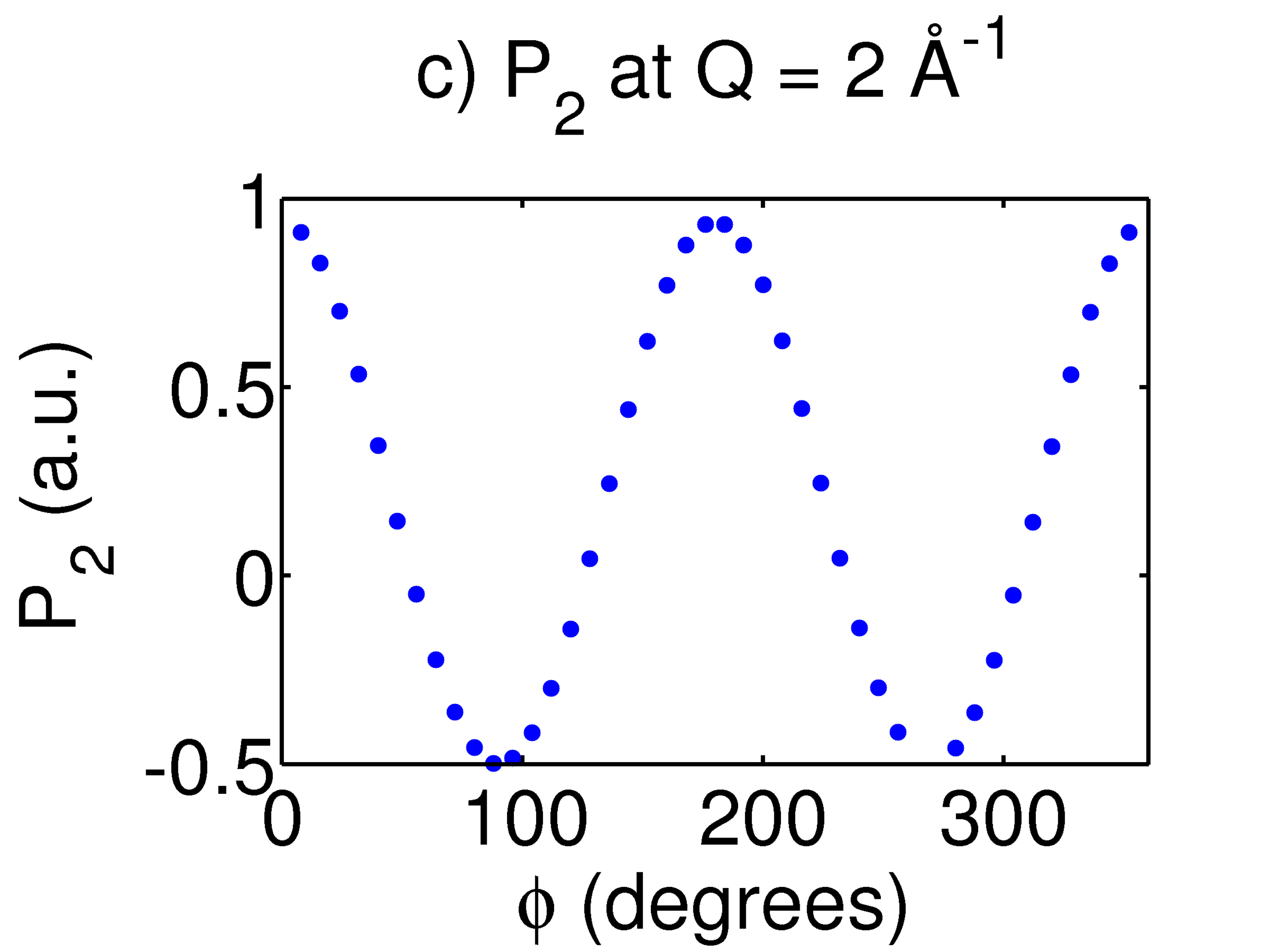}
\includegraphics[scale=0.25,bb=0 0 800 600]{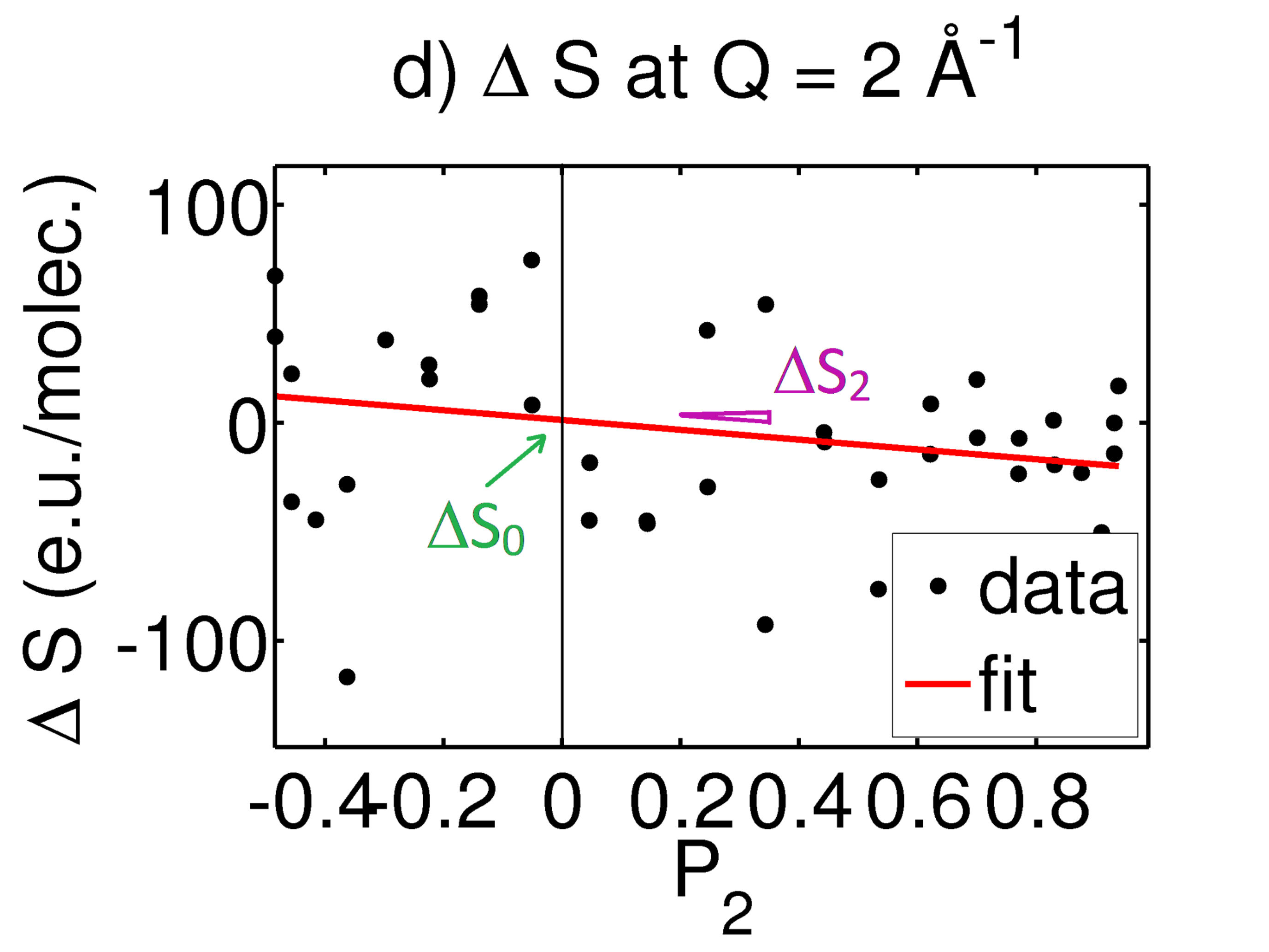}
}
\caption{\small \textbf{a)} $P_{2}(\cos\theta_{Q})$, with $\theta_{Q}$ as in Eq.~\ref{costhetaq_final}, mapped on the CSPAD. The azimuthal angle ($\varphi$) is defined to be zero at the projection of the laser polarization axis on the detector surface. A radial bin corresponding to \textit{Q} = 2 \AA{}$^{-1}$ is selected.  \textbf{b)} Averaged difference scattering pattern at 4.5 ps after photoexcitation of PtPOP in water. For comparison, a full scattering pattern is shown in Fig.~S1. \textbf{c)} Value of $P_{2}$ for the specific radial bin selected in panel (a) as a function of azimuthal angle $\varphi$, corresponding to \textit{Q} = 2 \AA{}$^{-1}$. \textbf{d)} Difference scattering signal $\Delta S$ at \textit{Q} = 2 \AA{}$^{-1}$ versus $P_{2}$ (black dots) and fitted straight line (red). The intercept with $P_{2}$ = 0 yields the isotropic scattering signal and the slope yields the anisotropic scattering signal, according to Eq.~\ref{allDiff}. The uncertainties ($\sigma_{\Delta S_{0}}$ and $\sigma_{\Delta S_{2}}$) are also estimated from the fit and further described in the text.}
\label{2dfigs}
\end{figure}

Fig.~\ref{2dfigs} illustrates the extraction of the isotropic and anisotropic difference scattering signals from a difference scattering pattern measured 4.5 ps after the photoexcitation of PtPOP in water. The scattering pattern is shown in Fig.~\ref{2dfigs}(b) and it is an average of 50 difference scattering patterns collected in a $\sim$ 40 fs wide time bin centred at 4.5 ps after the arrival of the laser pump pulse. As introduced above, such a signal arises mainly  from the photoinduced contraction of the Pt atoms along their connecting vector. In the excited state ensemble, this vector has a cosine-squared distribution with respect to the laser polarization axis, since the transition dipole moment is parallel to the Pt-Pt axis. Fig.~\ref{2dfigs}(a) shows $P_{2}$ [$P_{2} = \frac{1}{2}(3\cos^{2}\theta_{Q} - 1)$ and $\cos\theta_{Q}$ as in Eq.~\ref{costhetaq_final}] mapped onto the CSPAD, with $\varphi$ = 0 along the direction of the projection of the laser polarization axis on the detector surface, which was found to be inclined 20 degrees with respect to vertical for the experiment described here. 

\begin{figure}
\begin{center}
\includegraphics[scale=0.4,bb=0 0 800 400]{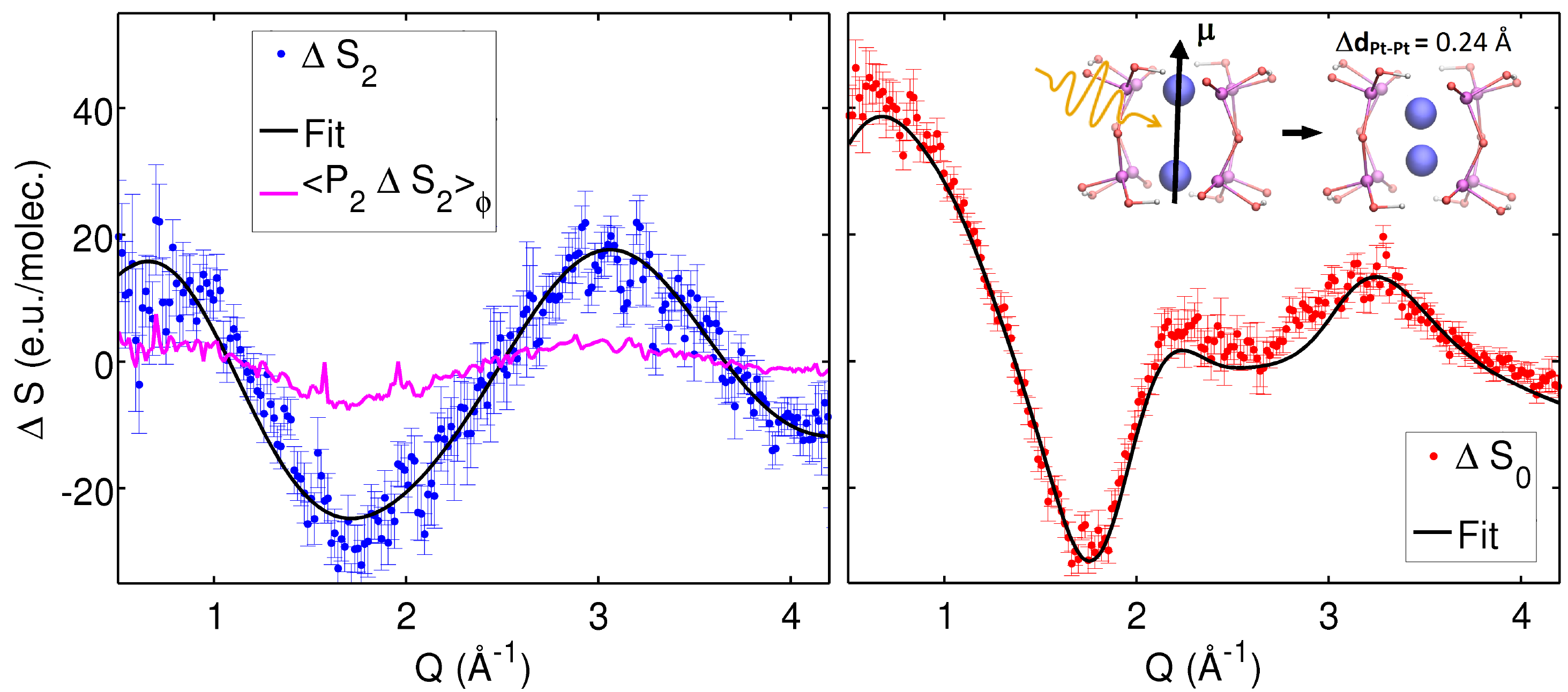}
\end{center}
\caption{ $\Delta S_{2}$ (left, blue points) and $\Delta S_{0}$ (right, red points) extracted from the pattern in Fig.~\ref{2dfigs}(b) through Eq.~\ref{allDiff}. The black line shows the simulated difference scattering signal through  Eq.~\ref{mains0} and Eq.~\ref{mains2} using as parameters the best-fit results obtained by a combined fit of  $\Delta S_{2}$ and $\Delta S_{0}$. The magenta line in the left panel shows the azimuthal average of the anisotropic signal ($\langle- P_{2} (\cos\theta_{Q}) \Delta S_{2}\rangle_{\varphi}$). }
\label{fighe1D}
\end{figure}

For the analysis, the scattering pattern in Fig.~\ref{2dfigs}(b) was divided in 500 radial bins and 45 azimuthal bins, and the signal in each bin was calculated as the average value of the pixels. The difference scattering signal corresponding to the radial bin centred at \textit{Q} = 2 \AA{}$^{-1}$ (black circle) is plotted in Fig.~\ref{2dfigs}(d) as a function of $P_{2}(\cos\theta_{Q})$ at the same \textit{Q} value. The red line is a least square fit of a straight line to the data points. This fit, according to Eq.~\ref{allDiff}, yields $\Delta S_{2}$ (\textit{Q} = 2 \AA{}$^{-1}$) as the slope and $\Delta S_{0}$ (\textit{Q} = 2 \AA{}$^{-1}$) as the intercept with the $P_{2}$ = 0 axis. Repeating the procedure for all the radial bins yields the one-dimensional isotropic and anisotropic difference scattering curves for the full \textit{Q} range at this specific time delay, as Fig.~\ref{fighe1D} shows.
 
\section{Results and discussion}
\begin{figure}
\begin{center}
\includegraphics[scale=0.4,bb=0 0 800 400]{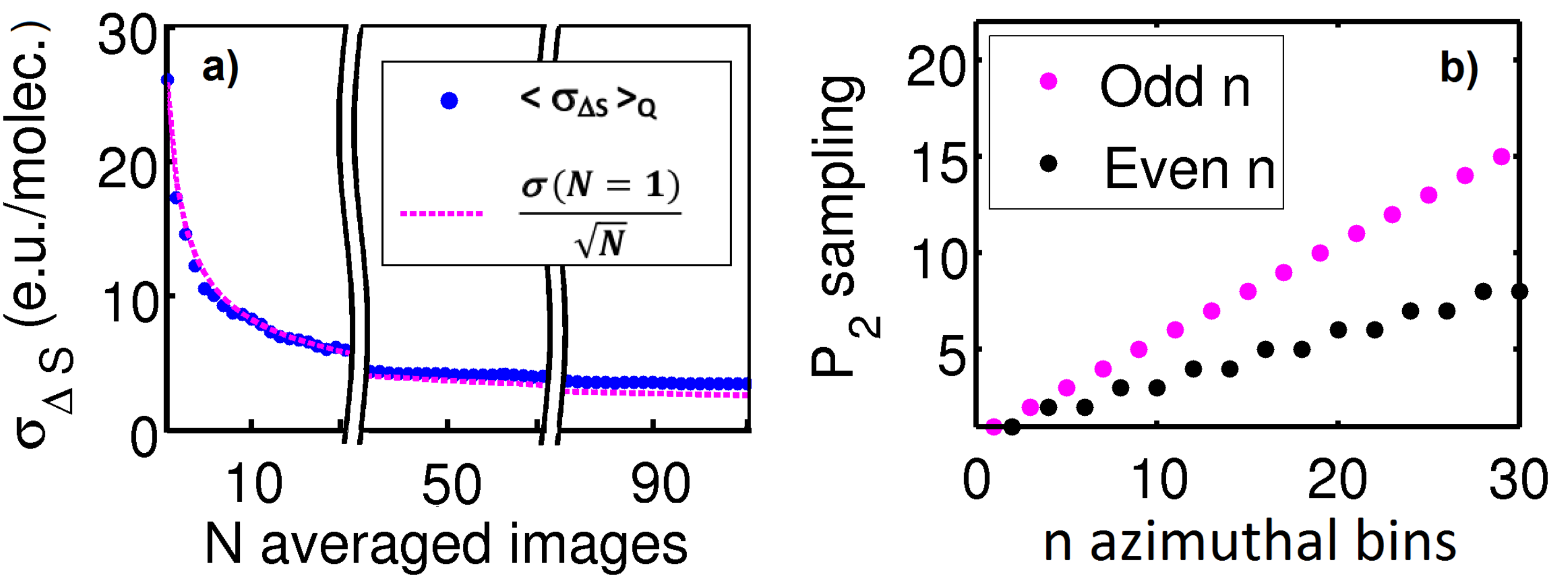} 
\end{center}
\caption{\small \textbf{a)} Average of the uncertainties $\sigma$(\textit{Q}) on $\Delta S_{2}$ over the full \textit{Q}-range (500 radial bins) as a function of number of images averaged in the time bin centred at 4.5 ps. The uncertainties are estimated from the covariance matrix of the straight line fit  (Eq.~\ref{allDiff}). The dashed line shows that the magnitude of the uncertainties up to $\sim$ 50 averaged images follows the behaviour of counting noise expected for a Gaussian distribution ($\sigma \propto N^{-1/2}$). \textbf{b)} Number of different $P_{2}$ values that are sampled as a function of number of azimuthal bins. This parameter is maximized if the number of azimuthal bins is odd.}
\label{methods1}
\end{figure}

Fig.~\ref{fighe1D} shows the isotropic and anisotropic difference scattering signals extracted from the scattering pattern measured at 4.5 ps  as a function of \textit{Q}. The uncertainty estimates for $\Delta S_{0}$ and $\Delta S_{2}$ at each \textit{Q} point are calculated from the covariance matrix of the coefficients of the straight line fit. Since the difference scattering signal is usually on the order of 0.1-1$\%$ of the total scattering signal (see comparison between Fig.~S1 and Fig.~\ref{2dfigs}(b) for this experiment), averaging several difference scattering patterns acquired at the same nominal time delay allows the improvement of the S/N ratio before the extraction of the isotropic and anisotropic contributions. Experimental parameters such as the magnitude of the differential scattering cross section and the X-ray photon flux, as well as the number of radial bins considered for each 2D pattern, determine the optimum number of images to be averaged for a specific experiment. The number of radial bins should be higher than the number of independent data points in the \textit{Q} range ($\sim$ tens of points~\cite{haldrup2010analysis}) and can be chosen arbitrarily high since re-binning can be done at a later stage of the analysis. For the present experiment, we set the number of radial bins to 500 and, for each time bin, we choose the number of images to be averaged by inspection of the decreasing uncertainties $\sigma$ on the coefficients of the straight line fit ($\Delta S_{0}$ and $\Delta S_{2}$ in Eq.~\ref{allDiff}) as a function of number of averaged images. This is exemplified in Fig.~\ref{methods1}(a) for the signal extracted at 4.5 ps: up to 50 averaged images $\sigma_{\Delta S}$ decreases as the inverse square root of the number of averaged images, as expected for Gaussian noise; after 50 images $\sigma_{\Delta S}$ converge to a constant value ($\sim$ 4 e.u./molec.) since the noise is then dominated by systematic errors due to contributions from a non-constant background and the non-linear response of the detector rather than counting statistics~\cite{timSVD}. Therefore averaging more than 50 images will not further decrease the uncertainties on the measured signal.

\begin{figure}
\begin{center}
\includegraphics[scale=0.45,bb=0 0 800 400]{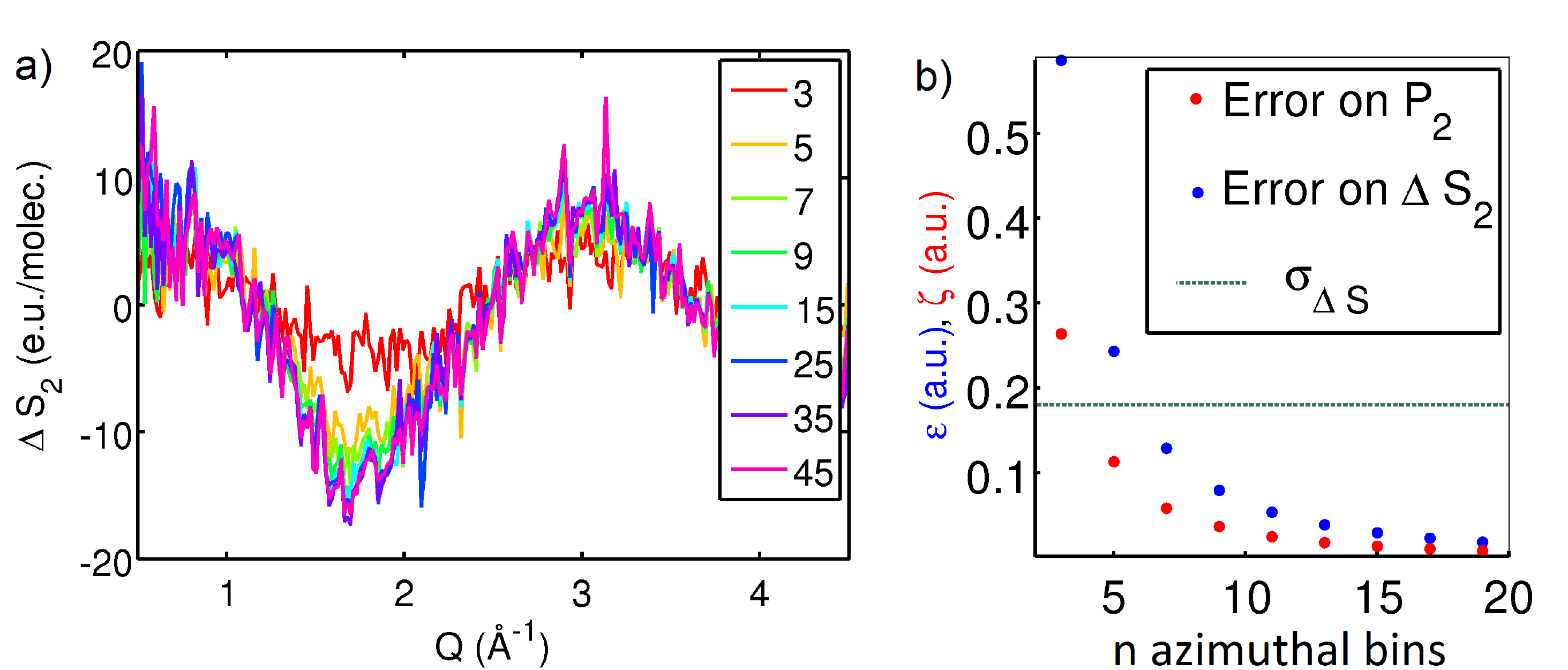} 
\end{center}
\caption{\small \textbf{a)} Anisotropic difference scattering signals extracted from the 2D scattering patterns in Fig.~\ref{2dfigs}(d) as a function of number of azimuthal bins. A significant decrease in the intensity of the  signal is observed when decreasing the number of azimuthal bins below 5. \textbf{b)} The red dots shows the difference ($\zeta$) between $P_{2}$ calculated as central value in the azimuthal bins or as in Eq.~\ref{asb}: $\zeta = \langle P_{2}(\cos\theta_{Q})\rangle_{2\delta\varphi} - P_{2}(\cos\theta_{Q}) = \frac{3}{4}\cos^{2}\theta\cos2\varphi(1-\frac{\sin2\varphi}{2\delta\varphi})$. With respect to this, the blue dots shows the resulting difference ($\epsilon$) in the magnitude of $\Delta S_{2}$ (at \textit{Q} = 2 \AA{}$^{-1}$): $\epsilon = (\langle\Delta S_{2}\rangle_{2\delta\varphi} - \Delta S_{2})/\langle\Delta S_{2}\rangle_{2\delta\varphi}$.The dashed black line shows the averaged experimental uncertainty $\sigma_{\Delta S_{2}}$ (divided by the measured signal). }
\label{methods2}
\end{figure}

Fig.~\ref{methods1}(b) shows the number of unique $P_{2}$ values that are sampled for a specific \textit{Q} value as a function of the number of azimuthal bins which the 2D diffraction patterns is divided into. The plot shows that choosing an odd number of azimuthal bins maximizes the number of unique points available for the fitting of $\Delta S$ versus $P_{2}$. This behaviour arises from the dependency of $P_{2}$ on the azimuthal angle $\varphi$, as shown in Fig.~\ref{2dfigs}(c), and from choosing equi-angular azimuthal bins. Moreover, since $P_{2}$ is non-linear as a function of $\varphi$, we introduce an error by calculating $P_{2}$ as the central value rather than the mean value in each azimuthal bin. This error can cause a significant decrease of the magnitude of the anisotropic difference scattering signal when decreasing the number of azimuthal bins, as shown in Fig.~\ref{methods2}(a). In order to calculate the mean value of $P_{2}$, we should use the following expression:
\begin{equation}
\langle P_{2}(\cos\theta_{Q})\rangle_{2\delta\varphi} = \frac{1}{2\delta\varphi}\int_{\varphi - \delta\varphi}^{\varphi + \delta\varphi} P_{2}(\cos\theta, \cos\varphi) d\varphi  = \frac{3}{4} \cos^{2}\theta (1-\frac{\sin2\delta\varphi}{\delta\varphi}\cos2\varphi) -\frac{1}{2}
\label{asb}
\end{equation} 
where $\delta\varphi$ is half the size of the azimuthal bins. Fig.~\ref{methods2}(b) shows the difference between the values of $P_{2}$ (red dots) and of the magnitude of $\Delta S_{2}$ (blue dots) obtained when calculating $P_{2}$ either as the central or the mean value of the azimuthal bins. From this observation, if the 2D scattering pattern is divided in more than 5 azimuthal bins, such a difference is within the uncertainty of the measurements ($\sigma_{\Delta S}$, the dashed green line) and can therefore be neglected, thus simplifying the calculations.

Since the isotropic difference scattering signal arises from the changes in radial distribution functions (RDFs) of the sample~\cite{asmusRDF}, $\Delta S_{0}$ in Fig.~\ref{fighe1D} comprises contributions arising from the changes in the structure of the solute and of the solvation shells, as well as from the changes in temperature and density of the bulk solvent~\cite{haldrup2010analysis,KasperSolvent,IheeReview2010,cammarata2006solvent}. The anisotropic scattering signal in Fig.~\ref{fighe1D} arises from the changes in the structure of the solute and potentially anisotropic changes in the solvation shell structure. The anisotropic response of the bulk solvent is discussed at the end of the section and, in case of water, its contributions are negligible at time scales longer than few hundred femtoseconds after excitation. In this experiment, the signal arising from the changes in the structure of the solute ($\Delta S^{solute}$) is strongly dominated by the contraction of the Pt-Pt bond due to the very electron-rich Pt atoms. As Fig.~S2 shows, in the limited Q-range available for this experiment, possible contributions to $\Delta S^{solute}$ arising from  intra-molecular changes other than the Pt-Pt bond contraction are found within the uncertainties of the measured signal. Based on these considerations, we construct a modelling framework where the photoinduced structural changes in the structure of the solute are parametrized through a single structural parameter, $\Delta d_{Pt-Pt}$, which describes the changes in Pt-Pt distance from the ground to the excited state of the molecule. Specifically, $\Delta S^{solute}$ is simulated from a set of structures derived by varying only the Pt-Pt distance of a Density Functional Theory (DFT) optimized ground state geometry of PtPOP. Details of the DFT calculations are provided in the SOI. The full model used to fit the difference scattering signal measured at 4.5 ps after the photoexcitation of PtPOP is:
\begin{equation}
\Delta S^{model}(Q,\Delta d_{Pt-Pt}) = \alpha\Delta S^{solute}(Q, \Delta d_{Pt-Pt}) + \beta\Delta S^{cage}(Q) + \Delta S^{solvent}(Q),
\label{model}
\end{equation}
where the first term on the right-hand side is $\alpha_{0} \Delta S_{0}^{solute}$, in case of the isotropic signal, and  $\alpha_{2} \Delta S_{2}^{solute}$, in case of the anisotropic one. $\Delta S_{0}^{solute}$ and $\Delta S_{2}^{solute}$ are  calculated using Eq.~\ref{mains0} and Eq.~\ref{mains2}, respectively, $\alpha_{0}$ and $\alpha_{2}$ represent the fraction of excited state molecules contributing to the isotropic and anisotropic scattering, respectively. $\Delta S^{cage}$ describes the changes in the structure of the solvation shells and $\beta$ is a scaling factor. For the isotropic case, $\Delta S^{cage}$ was calculated from the RDFs of the solute-solvent atom pairs~\cite{asmusRDF}, as further described in the SOI, while it was not necessary to include this term in the analysis of the anisotropic signal, given the good quality of the fit without the inclusion. Finally, $\Delta S^{solvent}$ describes the changes arising from the heating of the bulk water and its calculation is detailed in the SOI. We note that, since the anisotropic signal is insensitive to the isotropic changes of the bulk solvent, fewer degrees of freedom are used in the description of the anisotropic data compared to the isotropic contribution. Eq.~\ref{model} was fitted to the measured difference scattering signal within a standard  $\chi^{2}$  minimization framework~\cite{Jun_Ihee_2010}.  From a simultaneous fit of the isotropic and the anisotropic scattering signal, we find $\Delta d_{Pt-Pt}$ = 0.24 $\pm$ 0.04 \AA{}, $\alpha_{0}$ = 2.6 $\pm$ 0.2 \%, $\alpha_{2}$ = 1.5 $\pm$ 0.2 \%.  The black line in Fig.~\ref{fighe1D} shows the model constructed from the best-fit results, which well-describe the data ($\chi^{2}$ = 1.9). $\Delta d_{Pt-Pt}$ is found in agreement with previous studies~\cite{PtPOP2009}; the 1-sigma confidence interval on this parameter (0.04 \AA{}) is found to be smaller than what obtained from fitting the isotropic and the anisotropic signals separately  (which gives 1-sigma confidence interval of 0.05 \AA{} and 0.09 \AA{}, respectively). Since changes in population fraction are not expected on these single-ps time-scales~\cite{GRAY2017297} and taking into account the decay of the anisotropic signal, as described below, we find that following the excitation event $\sim$ 2.6 \% of the molecules have been photoexcited and contribute to the isotropic signal, and approximately 1.8 \%  contribute to the anisotropic scattering. We interpret this difference as an indication of multi-photon excitation of PtPOP, where the multi-photon excitation takes place through transitions where the dipole moment is not parallel to the Pt-Pt axis~\cite{Stiegman1987}. This combined analysis may further allow the disentanglement of the contributions to the difference scattering signal arising from structural changes parallel or perpendicular with respect to the transition dipole moment. Specifically to PtPOP, the structural changes contributing to $\Delta S_{2}$ depend directly on their displacement with respect to the Pt-Pt axis ($\xi_{ij}$ in Fig.~\ref{simTop}): Eq.~\ref{mains2} is maximum (and positve) when the $\xi_{ij}$ = 0 degrees (i.e. when the structural change is displaced parallel to Pt-Pt axis) and minimum (and negative) when $\xi_{ij}$ = 90 degrees (i.e. when the structural changes occur in the plane perpendicular to the Pt-Pt axis). However, as detailed above, the difference scattering signal is mainly dominated by the Pt-Pt contraction, and this hinders the investigation of additional structural parameters in the analysis described here.  

The magenta line in Fig.~\ref{fighe1D} shows the signal obtained from an azimuthal integration of the anisotropic contribution re-projected on the detector surface ($\langle- P_{2} (\cos\theta_{Q}) \Delta S_{2}\rangle_{\varphi}$). According to Eq.~\ref{allDiff}, this trace corresponds to the difference between the difference scattering signal obtained from an azimuthal integration of the 2D scattering pattern and the isotropic signal ($\langle\Delta S$ - $\Delta S_{0}\rangle_{\varphi}$). For the PtPOP data, such a difference is found to be $\sim$ 20 \% of the magnitude of the isotropic contribution at time delays immediately after the excitation event, where the anisotropy is most pronounced. These observations show that the azimuthally integrated signal, at time scales shorter than the rotational correlation time of the solute, may not be a good approximation of the scattering arising from an isotropic ensemble of photoexcited molecules. This implies that at such time scales, the use of azimuthally integrated scattering signals in subsequent analysis should be justified by comparing the S$_{0}$ and azimuthally integrated signals.

\begin{figure}
\begin{center}
\includegraphics[scale=0.4,bb=0 0 800 500]{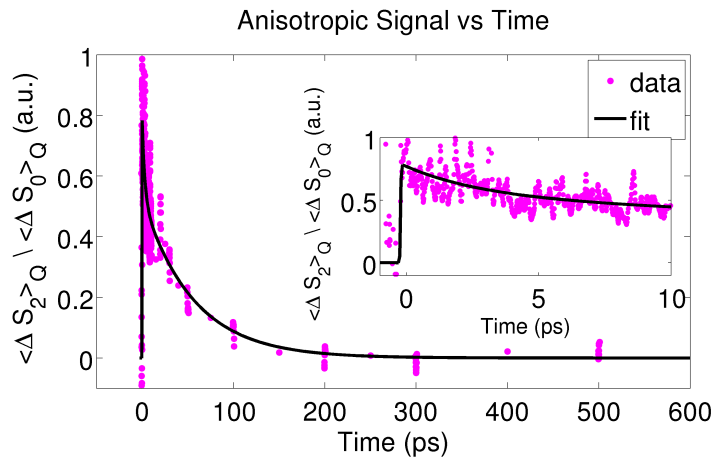}
\end{center}
\caption{\small Ratio between the magnitudes (calculated as averages of the absolute values of the signal in the full $Q$-range) of the anisotropic and isotropic difference scattering signals as a function of pump-probe time-delay, up to 600 ps after the photoexcitation of PtPOP in water. A broadened bi-exponential decay is fit to the data (black line), with time constants of 3 ps $\pm$ 2 ps and 60 ps $\pm$ 10 ps. The latter is interpreted as the rotational correlation time of PtPOP in water. }
\label{ANIdecayFig}
\end{figure}

Following the procedure described above, the anisotropic difference scattering signals were extracted from measured difference scattering pattens up to 1 $\upmu$s after photoexcitation of PtPOP (see Movie S1).  Fig.~\ref{ANIdecayFig} shows the decay of the anisotropy as a function of increasing pump-probe time delay. We find that 20 \% of the anisotropic contribution to the signal decays with a time constant of 3 ps $\pm$ 2 ps; while 80 \% decays with a time constant of 60 ps $\pm$ 10 ps. The latter is interpreted as the rotational correlation time of the PtPOP molecule in water. 
Using the Stokes-Einstein-Debye model~\cite{HorngRotationalDynamics1997}, the rotational correlation time of a sphere can be described as:
\begin{equation}
\tau_{r} = \frac{4 \pi \eta r^{3}}{3 k_{B} T}
\end{equation}
where $\eta$ is the viscosity of water, $k_{B}$ the Boltzmann constant, $T$ the temperature and $r$ the radius of the sphere. Approximating the PtPOP molecule to a sphere, and taking $r$ as half the longest inter-atomic distance in the PtPOP molecule ($r$ = 4 \AA{}), the rotational correlation time at room temperature can be estimated as 50 ps. This value is in agreement with the $\sim$ 60 ps found in this analysis. 

Finally, we note that the neat solvent may exhibit anisotropic scattering independently of the photoexcited solute molecules: the laser pulse weakly perturbs the equilibrium structure of the solvent through non-resonant excitations predominantly aligned with the polarization of the laser and the X-ray pulse probes the induced structural dynamics. Optically, the sample becomes birefringent, with different indices of refraction for light polarized parallel or perpendicular to the laser polarization axis. From an X-ray point of view, the scattering patterns are  observed to be anisotropic and the analysis described above can be applied to extract the anisotropic scattering as a function of  Q and time-delay. The time-evolution of the anisotropic X-ray scattering signal can be directly compared to the impulsive nuclear-coordinate response measured through techniques such as Raman-induced Kerr effect spectroscopy~\cite{KerrEffect,KerrEffect1995}. Since the time scales of this nuclear response response are known from optical Kerr effect studies, the anisotropic solvent scattering signal can be used to estimate the time-zero (the arrival time of the laser pulse at the sample) and the Instrument Response Function (IRF) of the experiment~\cite{Biasin2016}. The separate determination of these parameters significantly improves the analysis of the isotropic part of the scattering signal, by reducing the number of free parameters in the model used to fit the data~\cite{Biasin2016}. This method of extracting the IRF and time-zero can be used as a diagnostic for any time-resolved XDS experiment on solvated molecules where the anisotropic solvent response can be clearly identified.

\section{Conclusions}
In summary, the formalism introduced by Baskin and Zewail~\cite{baskin&zewaill_2006} and expanded by Lorenz $et$ $al$.~\cite{aniUlf} has here been applied to separate isotropic and anisotropic contributions from 2D difference scattering patterns measured upon photoexcitation of the solvated PtPOP molecule at an XFEL. The presented formalism has been directly implemented in quantitative structural analysis and we find that a combined analysis of the isotropic and anisotropic difference scattering signals helps the disentanglement of the many degrees of freedom involved in the structural response of the sample to photoexcitation and enhances the structural sensitivity. This analysis approach is generally applicable for all molecular systems with a well-defined transition dipole moment and asymmetric structural response to photoexcitation, provided a time-resolution shorter than the rotational dephasing time. Furthermore, we have discussed how the quantitative analysis of the anisotropic scattering signal can provide access to anisotropic solvent dynamics induced by the linearly polarized pump pulse and how these measurements can lead to independent determination of time-zero and the IRF of the experiment.  In conclusion, we have demonstrated a method that allows quantitative interpretation of anisotropic scattering signals measured at XFELs from aligned ensembles of molecules.  The information delivered by this method can benefit the overall interpretation of high-content scattering data from solution-state molecular systems. 


\ack{Acknowledgements}

The authors would like to acknowledge the beamline scientists of the XPP Instrument at LCLS. The DTU-affiliated authors would like to gratefully acknowledge DANSCATT for funding the beam time efforts. MMN, KBM, EB, MGL and AOD gratefully acknowledge support from the Danish Council For Indipendent Research under grant nr. DFF 4002-00272B. Use of the Linac Coherent Light Source (LCLS), SLAC National Accelerator Laboratory, is supported by the U.S. Department of Energy, Office of Science, Office of Basic Energy Sciences under Contract No. DE-AC02-76SF00515.

\referencelist[referencesJSR]

\end{document}